\documentclass[preprint,12pt]{aastex}       

\newcommand{\chisq}{$\chi^2$}
\newcommand{\civ}{C\,{\sc iv}}

\newcommand{\hi}{H\,{\sc i}}

\newcommand{\ciii}{C\,{\sc iii}]}
\newcommand{\cii}{C\,{\sc ii}]}

\newcommand{\feii}{Fe\,{\sc ii}}

\newcommand{\lya}{Ly$\alpha$}

\newcommand{\mgii}{Mg\,{\sc ii}}
\newcommand{\znii}{Zn\,{\sc ii}}
\newcommand{\siiv}{Si\,{\sc iv}}
\newcommand{\nv}{N\,{\sc v}}
\newcommand{\bump}{2175~\AA}
\newcommand{\etal}{et~al.}

\begin{document}

\title{Detections of the 2175~\AA\ Dust Feature at $1.4 \lesssim z \lesssim 1.5$ from the Sloan Digital Sky Survey}

\author{
Junfeng Wang,\altaffilmark{1} Patrick B. Hall,\altaffilmark{2}
Jian Ge,\altaffilmark{1} Aigen Li,\altaffilmark{3} and Donald P.
Schneider\altaffilmark{1}} \altaffiltext{1}{Department of
Astronomy \& Astrophysics, The Pennsylvania State University, 525
Davey Lab, University Park, PA 16802; jwang@astro.psu.edu, jian@astro.psu.edu, dps@astro.psu.edu} \altaffiltext{2}{Princeton
University Observatory, Princeton, NJ08544-1001; pathall@astro.princeton.edu}\altaffiltext{3}{Steward Observatory, and Lunar and
Planetary Laboratory, University of Arizona, Tucson, AZ 85721; agli@lpl.arizona.edu}

\begin{abstract}
The strongest spectroscopic dust extinction feature in the Milky
Way, the broad absorption bump at 2175~\AA, is generally believed
to be caused by aromatic carbonaceous materials -- very likely a
mixture of Polycyclic Aromatic Hydrocarbon (PAH) molecules, the
most abundant and widespread organic molecules in the Milky Way
galaxy. In this paper we report identifications of this absorption
feature in three galaxies at $1.4 \lesssim z \lesssim 1.5$ which
produce intervening \mgii\ absorption toward quasars discovered by
the Sloan Digital Sky Survey (SDSS). The observed spectra can be fit
using Galactic-type extinction laws, characterized by parameters
[$R_V$, $E(B-V)$] $\simeq$ [0.7, 0.14], [1.9, 0.13], and [5.5,
0.23], respectively, where $R_V\equiv A_V/E(B-V)$ is the
total-to-selective extinction ratio, $E(B-V)\equiv A_B-A_V$ is the
color-excess. These
discoveries imply that the dust in these distant quasar absorption
systems is similar in composition to that of Milky Way, but with a
range of different grain size distributions. The presence of
complex aromatic hydrocarbon molecules in such distant galaxies is
important for both astrophysical and astrobiological
investigations.
\end{abstract}

\section{Introduction}
The space between the stars of the Galaxy and external galaxies is
filled with gaseous ions, atoms, molecules and tiny dust grains.
These interstellar grains, spanning a wide range of sizes from a
few angstroms to a few micrometers, are important for the
evolution of galaxies, the formation of stars and planetary
systems. So far, one of the most best-studied properties of
interstellar dust is its obscuration of starlight. The size and
composition of interstellar dust are mainly inferred from,
respectively, the extinction curve and its spectral features. The
strongest spectroscopic interstellar extinction feature in the
Galaxy is the broad \bump\ absorption bump. This feature was first
detected by the Aerobee rocket observations \citep{St65}. In this
work we report the detection of the 2175~\AA\ dust exinction bump
from individual intervening absorption systems in the spectra of
three Sloan Digital Sky Survey \citep[SDSS;][]{York00} quasars.

The review of \citet{dr03} provides an excellent summary of the
properties of the 2175~\AA\ dust extinction bump and theoretical
constraints on its carrier. This feature is seen in extinction
curves along lines of sight in the Milky Way (MW), the Large
Magellanic Cloud (LMC), and some regions of the Small Magellanic
Cloud (SMC). Extinction curves in the SMC bar region lack the
2175~\AA\ feature. The central wavelength of the feature varies by
only $\pm0.46\%~(2\sigma)$ around 2175~\AA, while its FWHM varies
by $\pm12\%~(2\sigma)$ around 469~\AA. Given its substantial band
strength, the carrier responsible for the feature must contain at
least one of the most abundant elements C, O, Mg, Si or Fe.
Although the exact carrier is unknown, Draine (2003) states that
``it now seems likely that some form of graphitic carbon is
responsible'', most likely polycyclic aromatic hydrocarbons (PAHs; Joblin, Leger, \& Martin 1992; Li \& Draine 2001; Draine 2003 and references within).

Several previous detections have been reported of the observed
\bump\ feature in distant galaxies using quasar spectra.
\citet{Pit00} presented a good summary of the state of the field
as of three years ago, and ruled out several reported detections,
which will not be discussed here.

The highest redshift \bump\ bump observation to date is that of
\citet{Vernet01}, who claimed a probable detection in a composite spectrum of
radio galaxies at $z\sim2.5$, although they acknowledged that the detection is
difficult to confirm due to the low signal-to-noise ratio (SNR) and blending
with \feii\ emission at $\lambda>2300$~\AA.  Additionally, the dust responsible
for the feature does not appear to be located in the host galaxies proper,
but rather in the narrow emisson-line regions of these radio galaxies
where the physical conditions may be quite different,
perhaps leading to different carriers for the bump.

The only previous detection of this feature from an individual
intervening absorption system was that of \citet{Cohen99}, who
detected the \bump\ feature in a damped Ly$\alpha$ absorber at
redshift $z=0.524$ toward the BL Lac object AO~0235+164 at
$z=0.94$. They found a lower dust-to-gas ratio than in the Galaxy
and mentioned that the \bump\ feature ``suggests there are
differences from the average Galactic curve.''\footnote{The
average Galactic curve is parameterized by $R_V=3.1$, where
$R_V\equiv A_V/E(B-V)$ is the ratio of total to selective
extinction and $E(B-V)\equiv A(B)-A(V)$ is the color excess.}

Such differences appear to be rather common when the \bump\ feature is detected
with data of sufficient quality to model the extinction curve involved.
\citet{Ma97} detected the \bump\ feature in the composite absorption spectrum
of 96 intervening \mgii\ absorption systems at redshifts $0.2 < z < 2.2$.
The strength of the feature was roughly consistent
with a standard Galactic dust-to-gas ratio.
\citet{Falco99} reported detections in several $z \lesssim 1$
galaxies responsible for the gravitational lensing of background
quasars. Many of the estimated extinction curves do not match the
standard $R_V=3.1$ Galactic curve (e.g., $R_V=7.2\pm 0.1$ was
found for a $z_l=0.68$ spiral galaxy).
\citet{Toft00} detected the \bump\ feature from the lensing galaxy of B~1152+199
at $z=0.44$ and fitted its extinction curve by a Galactic-type extinction law
with $1.3\lesssim R_V \lesssim 2.1$ and $0.9 \lesssim E(B-V) \lesssim 1.1$.
\citet{Motta02} detected a strong \bump\ bump in a lensing galaxy
at $z=0.83$. Their data are well fitted by a standard $R_V=2.1\pm
0.9$ Galactic extinction curve, leading them to suggest that the
lens galaxy of SBS 0909+532 contains dust like that of the Galaxy.
\citet{Wuck03} may have marginally detected the \bump\ feature in
a lensing galaxy (later identified with a damped Ly$\alpha$
absorber) at $z=0.93$. They found that using extinction curves
with a significant 2175~\AA\ bump reproduces the data better than
curves without this feature.
Recently \citet{mu04} reported that the dust in the $z_l=0.68$ lens galaxy of
B~0218+357 shows the \bump\ bump but
produces a very flat ultraviolet extinction curve with $R_V=12\pm 2$.

Given the range of extinction curves found in these extragalactic
systems, large and well-selected samples extending to higher
redshifts are needed to characterize the diversity and evolution
of dust properties in the early universe. As a step towards this
goal, we present the \bump\ feature identified in three individual
intervening absorption systems in the spectra of three SDSS
quasars.

\section{Data}

One of the goals of the SDSS is to obtain spectra for $\sim$10$^5$ quasars to $i=19.1$ ($i=20.2$ for $z>3$
candidates), in addition to the $\sim10^6$ galaxies that comprise the bulk
of the spectroscopic targets \citep{St02,Bl03}.
From astrometrically calibrated drift-scanned imaging data \citep{Gu98,Pi03}
on the SDSS $ugriz$ system \citep{Fu96}, quasar candidates are selected primarily using color criteria designed to target objects whose broad-band colors are different from those of normal stars and galaxies (Richards et~al. 2002). Due to these inclusive criteria, the selection of candidates using $i$ band
magnitudes rather than blue magnitudes which are more affected by absorption
and reddening, and its area and depth, the SDSS can reveal
quasars with previously rarely seen properties.  Relevant instrumental and observational details of the SDSS Data
Release One (hereafter DR1) can be found in \citet{Ab03}. The DR1
spectra distributed by the SDSS cover approximately
3800--9200\,\AA\ and have been sky subtracted, wavelength
calibrated to the heliocentric frame, corrected for telluric
absorption and also for Galactic extinction. One of us (P. B. H.)
carried out a visual inspection of the SDSS spectra of the $\simeq
6350$ DR1 quasars with $z \geq 1.6$ (Schneider et~al. 2003). The
goal was to identify examples of unusual quasar subtypes
(primarily broad absorption line quasars, nitrogen-strong quasars,
and dust-reddened quasars) for later comparison with the results
of automated searches. This inspection uncovered several quasars
with a possible \bump\ signature, the most convincing three of
which are SDSS J145907.19+002401.2, SDSS J144612.98+035154.4 and
SDSS J012147.73+002718.7 (hereafter simply SDSS J1459+0024, SDSS
J1446+0351 and SDSS J0121+0027).

\section{Fitting and Results}
The spectral index $\alpha$ redward of the Lyman $\alpha$ emission
line --- defined by fitting a power-law $f_{\lambda} \propto
\lambda ^ {-(\alpha+2)}$ to the continuum --- confirms possible
reddening in these quasar spectra when compared to the composite
quasar spectrum generated using a homogeneous data set of more
than 2200 SDSS quasar spectra (Vanden Berk \etal\ 2001). As shown
in Figure 1, compared to that composite's mean spectral index of
$\approx -0.46$, SDSS J1446+0351 has $\alpha \approx -1.93$, SDSS
J1459+0024 has $\alpha \approx -1.91$, and SDSS J0121+0027 has
$\alpha \approx -1.27$, all suggesting either unusally red
intrinsic continua or heavily dust-reddened spectra. We
synthesized $g$, $r$, and $i$ magnitudes from each spectrum independently and
compared these values to the SDSS photometric measurements; this
comparison demonstrated that the spectrophotometric calibrations
of the spectra are accurate.

Since quasars have a range of intrinsic continuum slopes, to
investigate the dust reddening properties in these absorption
systems we simulate the dust-reddened quasar spectrum using
template quasar spectra. The templates include the average SDSS
quasar composite spectrum from \citet{Va01}, and the
reddest-quartile and bluest-quartile composites from \citet{Ri03}
(composite spectra of SDSS quasars binned by continuum color). We
adopt different types of extinction laws and treat $E(B-V)$, the
amount of extinction, as a free parameter. We adopt the
$E(\lambda-V)/E(B-V)$ relations from Pei (1992) to calculate
empirical extinction curves for the Milky Way, LMC and SMC.  We
also consider the ``CCM Galactic extinction law'' (Cardelli,
Clayton \& Mathis 1989), which gives a parameterized analytical
form of $A(\lambda)/A(V)$ for wavelengths $\lambda >$ 1000~\AA:
\[\frac{A(\lambda)}{A(V)}=a(x)+\frac{b(x)}{R_V}\]
where $a(x)$ and $b(x)$ are uniquely defined curves as a function
of the wavenumber $x=\lambda ^{-1}$, and $R_V \equiv A(V)/E(B-V)$.
(Note that the Milky Way empirical extinction curve used in Pei
(1992) is equivalent to the CCM extinction law with $R_V=3.08$,
the mean value from measurements along different lines of sight.)
These extinction curves are then applied to redden the template
spectrum.

The simulated spectra are normalized to the data using the mean
flux density between 7000--7500\,\AA\ or 7500--8000\,\AA\ in the
observed frame, since the red end of each spectrum is least
affected by dust extinction. The reduced $\chi^2$ is calculated
for each fit using the entire spectrum except for the emission
line wavelength regions as defined in Vanden Berk \etal\ (2001)
and for the wavelengths of narrow absorption lines.

We now discuss the fitting results for the three individual
quasars.  Table 1 lists the redshifts of the quasars and of all
absorption lines identified in the intervening systems associated
with the putative 2175\,\AA\ bumps. The narrow absorption line
parameters were measured using the IRAF\footnote{IRAF is
distributed by the National Optical Astronomy Observatories, which
are operated by the Association of Universities for Research in
Astronomy, Inc., under cooperative agreement with the National
Science Foundation.} standard software package {\it splot}. We
assume that all dust along the line of sight is located at these
intervening redshifts.  For clarity, all the fitting attempts with
different quasar composite spectra are summarized in Table 2; all
the \bump\ measurements and the metallicity estimates for three
absorption systems are summarized in Table 3.

\subsection{SDSS J1459+0024}
The quasar redshift is $z =3.0124\pm 0.0005$, and a strong
intervening absorption system is seen at $z =1.3946\pm 0.0001$
(all absorption redshifts are measured from the \mgii\ doublet).
The broad absorption feature centered between the \lya\ 1215 and
\civ\ 1550 emission lines is consistent with being the
 \bump\ extinction feature from dust in the intervening
absorption system at $z=1.3946$. The restframe width of the broad absorption
feature is $\sim$ 400$\pm60$~\AA\ FWHM, and its central wavelength is 2230$\pm20$~\AA\
at a redshift of $z = 1.395$. The measurements of FWHM and central wavelength are obtained with a Drude-like profile fit of the observed absorption feature\footnote{The CCM functional form used a Drude-like profile to represent the \bump\ extinction hump: $\sim x^2/[(x^2-x_0^2)^2+\gamma ^2 x^2]$, $x\equiv \lambda^{-1}$, $x_0=\lambda_0^{-1}$, where $\lambda_0$ is the central wavelength and $\gamma$ is the FWHM. $x_0$ and $\gamma$ are determined by fitting the observed spectrum \citep{Fi86,ccm89}.} (the same measurements for $\S 3.2$ and $\S 3.3$). The 1$\sigma$ error is estimated by including different regions of continuum bracketing the feature into the fit (i.e., fit for several wavelength ranges between 2050--2550~\AA, in the dust restframe.). In the Milky Way this feature is
observed to have an almost constant central wavelength
$\lambda_0=2175 \pm 9$~\AA. This central wavelength difference
between the Milky Way and this absorption system likely arises
from the uncertainty caused by the \nv\ 1240 and \siiv\ 1398
emission lines bracketing the absorption feature. We rule out identifying this feature as a \civ\ broad absorption line trough
because there is no evidence for broad absorption in other lines commonly
seen in BAL quasars (e.g., \nv).

By varying $E(B-V)$ with the Milky Way, LMC and SMC empirical
curves, it is found that the best fit to both the broad absorption
feature and the large scale reddening is obtained using a Milky
Way-type extinction curve with $E(B-V) = 0.2$.  The \bump\ bump in
the simulated spectra reddened with the LMC or SMC curves is not
sufficiently deep to match the bump in the observed spectrum,
indicating the existence of Milky Way-type dust in this absorber.
To find a better fit to the flux level and the overall shape of
the broad absorption, we add in $R_V$ as another free parameter by
adopting the CCM extinction law. With the CCM extinction law, the
best fit is obtained with the reddest composite, which gives
$E(B-V) = 0.13\pm 0.01$, $R_V = 1.9^{+0.3}_{-0.2}$ and a minimum
reduced $\chi^2 \sim 1.8$. The 1$\sigma$ error bar is estimated by
variation of \chisq. With the average and the bluest composite, we
can still obtain a good overall fitting with slightly larger
\chisq\ (Table 2, Fig. 2). The fitting with composite spectra of
different colors indicates $0.7\lesssim R_V \lesssim 1.9$; the
rather small value of $R_V$ compared to the average value of 3.1
in the Milky Way \citep{Sa79} indicates that the size distribution
of the dust grains is skewed towards substantially finer
particles.\footnote{Although the exact $R_V$ value is dependent on
the adopted composite spectrum type (reddest, average, or bluest),
it is somewhat secure to conclude that this absorber system favors
smaller $R_V$ (than that of the Milky Way) and the SMC-type
extinction curve without the \bump\ hump is not able to give a
good fit.}

Note that SDSS J1459+0024 is the only one of our three quasars
detected by 2MASS, and its $J-K$ color of $1.22\pm 0.17$ places it
in the reddest quartile of $z=3$ quasars \citep{Ba01}. The dust in
the intervening galaxy would not have much of an effect at the
rest-frame wavelengths probed by the $J$ and $K$ bands, so the red
$J-K$ color of SDSS J1459+0024 is good evidence that its
UV-optical continuum is intrinsically red. This is consistent with
our result that the best fit is obtained with the reddest quartile
composite.

\subsection{SDSS J1446+0351}
The quasar redshift is $z = 1.9452\pm 0.0023$, and a strong
intervening absorption system is seen at $z =1.5115\pm 0.0002$.
Only a few absorption lines are identified in this absorption
system, as listed in Table 1. The quasar spectrum shows a broad
absorption feature centered between the \civ\ 1550 and \cii\ 2326
emission lines, consistent with the \bump\ extinction feature from
dust in the intervening absorption system at $z = 1.51$.  The
width of the broad absorption feature is $\sim$ 450$\pm55$~\AA\ FWHM, and its
central wavelength is 2200$\pm15$~\AA\ at a redshift of $z = 1.51$. This
small wavelength shift is probably due to the uncertainty produced
by the \ciii\ 1909 line superposed on the absorption feature.

With $E(B-V)=0.28$ we obtain the best fit of the broad absorption
feature and the large scale reddening with the empirical Milky
Way-type extinction curve. Again, the \bump\ absorption in the
simulated spectra based on LMC and SMC curves is not deep enough
to match the observations. With the CCM extinction law, we obtain
the best fit using the reddest composite with $E(B-V) =
0.14\pm0.01$, $R_V = 0.7^{+0.4}_{-0.2}$ and a minimum reduced
$\chi^2 \sim 2.6$. Using the average and the bluest composite, we
obtained almost identical curves to the best fit with reddest
composite and $0.5\lesssim R_V \lesssim 0.7$ (Table 2, Fig. 3).
This implies our fitting is not sensitive to the intrinsic quasar
spectrum and the unexpected small value of $R_V$ is not an
artifact. The dust grain size distribution in this system also
appears to be skewed towards smaller particles.

\subsection{SDSS J0121+0027}
The quasar redshift is $z = 2.2241\pm 0.0001$, and a strong
intervening absorption system is seen at $z = 1.3880\pm 0.0001$.
Standard lines identified in this absorption system are also
listed in Table 1. Two other absorption systems are also seen in
the spectrum: \civ, \siiv\ and possibly weak \mgii\ absorption at
$z=1.955\pm 0.001$, and \civ, probably \nv\ and possibly
Ly$\alpha$ absorption at $z= 2.200\pm 0.005$. Again the quasar
spectrum shows a broad absorption feature, centered between the
\civ\ 1550~\AA\ and \ciii\ 1909~\AA\ emission lines, consistent
with being the \bump\ feature from dust in the $z=1.388$
intervening absorption system. The central wavelength in the rest
frame for the broad absorption feature is 2250$\pm25$~\AA\ and the
restframe width is $\sim$ 400$\pm60$~\AA\ FWHM at a redshift of $z =
1.388$.  The \civ\ emission probably causes the apparent shift of
the dust absorption feature toward longer wavelengths than seen in
the Milky Way.

Using the Vanden Berk (2001) SDSS composite spectrum and simply
varying $E(B-V)$ with the empirical curves, one cannot achieve a
satisfactory overall fit to this quasar's spectrum.  With the
bluest-quartile composite spectrum from \citet{Ri03} and $E(B-V) =
0.14$, we obtain the best fit of the broad absorption feature and
the large scale reddening using the Milky Way empirical
extinction. With the LMC and SMC empirical extinction curves, the
\bump\ absorption is not sufficient to match the observed
amplitude, also indicating Galactic-type dust content in this
absorber (Fig. 4).

We applied the CCM Galactic extinction law to the same
bluest-quartile composite.  We obtain the best fit with
$E(B-V)=0.23^{+0.02}_{-0.01}$, $R_V = 5.5\pm 0.1$ and a minimum
reduced $\chi^2 \sim 1.3$ (Fig. 4). Using the reddest composite of
\citet{Ri03} leads to difficulty matching the flux in the far-UV
region. We estimate a range of $5.5\lesssim R_V \lesssim 6.0$ from
the fitting results using the average and bluest composite (Table
2). The large value of $R_V$ for this system implies that large
dust grains are more prevalent than in the Milky Way diffuse ISM.

\subsection{Discussion}
The best fit parameters of both of the first two systems (SDSS
J1459+0024 and SDSS 1446+0351) include surprisingly small $R_V$
values, implying a relative overabundance of finer dust grains
compared to Milky Way dust.\footnote{The previously reported
minimum $R_V=1.5$ is associated with an elliptical galaxy at
$z=0.96$ \citep{Falco99}.} In the Milky Way, so far the most
extreme extinction curve is found for the line of sight toward the
HD 210121 high latitude translucent cloud. The extinction curve,
with $R_V\approx 2.1$, is characterized by an extremely steep
far-UV rise and by a relatively weak and broader \bump\ hump
\citep{We92, La96}. Detailed extinction, polarization and IR
emission modeling implies that this region is rich in finer dust
grains \citep{Li98, La00, We01, Cl03}. Detailed modeling of the
extinction curves of these intervening systems in terms of the
silicate/graphite-PAHs model \citep{Li031} are in progress (J.Wang
\etal\ 2004, in preparation). Our goal is to infer their dust size
distribution and composition. However, there are other possible
explanations. For example, if the quasars are intrinsically redder
than indicated by our fits, to reproduce their observed far-UV
spectra would require extinction curves which are less steep in
the far ultraviolet (i.e., which have higher $R_V$). We note that
there are also uncertainties in the 5--6 $\mu$m$^{-1}$ range of
the CCM extinction curve.

Nonetheless, since the physical conditions governing dust
properties evolve with redshift (e.g., star formation rate,
metallicity, etc.), it should not be surprising that the dust
properties also evolve with redshift. If the diversity of grain
sizes is real and not due to biases in the extinction modeling (in
both our work and that of other groups mentioned in $\S$1), it is
extremely interesting for understanding dust formation at high
redshift. Generally speaking, preferential removal of small dust
grains (e.g. by coagulation in dense molecular clouds) will result
in a gray extinction law with large $R_V$, while in star forming
galaxies finer grains will dominate the dust size distribution as
a result of dust destruction by grain-grain collisions or grain
shattering by shock waves.

Lack of the detailed knowledge of the metallicity of these
absorber systems, we use the Milky Way gas-to-dust ratios
$N$(\hi+H$_2$)/$E(B-V)=5.8\times 10^{21}$\,cm$^{-2}$\,mag$^{-1}$
(Bohlin, Savage, \& Drake 1978) and
$N$(\hi)/$E(B-V)=4.93\pm0.28\times 10^{21}$\,cm$^{-2}$\,mag$^{-1}$
(Diplas \& Savage 1994) to estimate the neutral gas content of
these absorption systems. From our best fit $E(B-V)$ values we
estimate $N$(\hi)$\approx 6.4\times 10^{20}$\,cm$^{-2}$ for SDSS
J1459+0024 and $N$(\hi)$\approx 6.9 \times 10^{20}$\,cm$^{-2}$ for
SDSS J1446+0351. The strength of the \mgii\ absorptions suggest
that those absorbers may be damped \lya\ systems, which have a
lower limit to their column densities of $N$(\hi)=$2\times
10^{20}$\,cm$^{-2}$. From the SDSS spectra of the two quasars
(spectral resolution $\sim$2000), the equivalent width ratio of
\mgii\ $\lambda$$\lambda$2796,2803 indicates that both lines are
either saturated or nearly so. Therefore we can constrain the
range of column densities for \mgii: $7.7\times 10^{13} \le$
$N$(\mgii)$\le 3.1\times 10^{17}$\,cm$^{-2}$ for SDSS J1459+0024
and $1.7\times 10^{14} \le$ $N$(\mgii) $\le 7.0\times
10^{17}$\,cm$^{-2}$ for SDSS J1446+0351. The lower and upper ends
of these ranges represent the linear and damping regimes of \mgii,
respectively. A lower limit estimate of the \mgii\ abundance
compared to solar abundance gives [Mg/H]=$-$2.7 for SDSS
J1459+0024 and [Mg/H]=$-$2.2 for SDSS J1446+0351. No \znii\ lines
were found in either system, so we cannot estimate the dust
depletion. Some recent surveys have discovered strong molecular
hydrogen lines in damped Ly$\alpha$ absorbers with strong dust
depletion, suggesting that the formation of H$_2$ on the dust
grains is the dominant formation process \citep{Ge97,Ge99,Ge01}.
HST observations of these quasars will be extremely valuable to
verify this hypothesis.

For SDSS J0121+0027, using the Milky Way gas-to-dust ratio we
estimate an average hydrogen column density $N$(\hi)$\approx
1.1\times 10^{21} $\,cm$^{-2}$ from our best fit $E(B-V)=0.23$.
Again, this suggests that the strong \mgii\ absorption system is a
damped Ly$\alpha$ absorption system.  Based on the equivalent
widths of the weak absorption lines of \znii\ 2026 and \feii\ 2587
in the $z= 1.388$ absorber, we estimate the column densities for
\znii\ and \feii\ are $3.4\times 10^{13} $\,cm$^{-2}$  and
$1.6\times 10^{14} $\,cm$^{-2}$, respectively, assuming they are
both on the linear part of the curve of growth. Since both \znii\
and \feii\ are the dominant species for these metals in DLAs, the
estimated Zn and Fe abundances are [Zn/H]\,$\simeq -0.2$ and
[Fe/H]\,$\simeq -2.6$, respectively. The relative depletion
between Fe and Zn, [Fe/Zn]\,$\sim -2.4$, is very large, similar to
heavily depleted diffuse clouds in the Milky Way, such as the
$\zeta$ Oph cloud ([Fe/Zn]\,$\simeq -1.6$, Savage \& Sembach
1996). This dust depletion is the largest among all of the
high-redshift DLAs searched for dust and molecular absorption to
date \citep{Ge01,Le03}. The large dust depletion provides
additional evidence that the strong dust extinction feature in
this system is real.

Although the precise nature of the carrier of the \bump\ extinction feature
remains unknown, it is generally accepted that this feature arises
in some types of aromatic carbonaceous materials \citep{dr03,Li03}.
A proposal involving a cosmic mixture of many individual PAH molecules, radicals, and ions is receiving increasing attention \citep{Li031}. PAHs, the most abundant and widespread organic molecules in the Milky Way Galaxy as well as in other nearby galaxies \citep{Al03}, reveal their existence in interstellar space by emitting a distinctive set of broad spectral lines at 3.3, 6.2, 7.7, 8.6 and 11.3 $\mu$m \citep{Li031}. PAHs are thought to form predominantly in the atmospheres of carbon stars \citep{La91, Allain96};
from there, stellar winds will deposit PAHs into the surrounding ISM.\footnote{We should note that the origin and evolution of interstellar PAHs are not very clear. Other suggested sources for interstellar PAHs include (1) shattering of carbonaceous interstellar dust or of photoprocessed interstellar dust organic mantles (Greenberg et al.\ 2000) by grain-grain collisions in interstellar shocks (Jones, Tielens, \& Hollenbach 1996); (2) {\it in-situ} formation through ion-molecule reactions (Herbst 1991). In the early universe, massive rate of star formation results in a much denser UV starlight intensity and a much greater supernova explosion rate. Both effects would enhance the destruction of PAHs. We would admit that even the origin and survival of PAHs in the Milky Way galaxy are not well understood. See Li (2004) for details.}
As the single largest repository of organic material in our galaxy,
PAHs play an important role in prebiotic chemistry which may ultimately lead
to the development of organic life. Under dense molecular cloud conditions,
energetic processing (starlight irradiation and/or cosmic ray bombardment) of
ices containing PAHs produces aromatic alcohols, ethers, ketones, and quinones;
these are essential for important processes in living systems
(Bernstein et~al. 1999, 2002, 2003).  Such complex organic molecules can be
delivered to the surfaces of planetary objects via dust particles \citep{Cl93}
and carbonaceous meteorites \citep{Ze89} --- the formation mechanism of complex
organic molecules found in such meteorites is not known for certain, but the
deuterium enrichment seen in many such molecules suggests an interstellar
origin \citep{Cr93}.

The presence of PAHs in distant galaxies may thus be of
astrobiological interest.  The galactic environments in the
galaxies discussed in this paper have probably evolved to the
stage where the above-mentioned organic molecules, and possibly
even molecules of ribonucleic acid and of various proteins, can
form on the surfaces of dust consisting of PAHs
\citep{Be021,Mu02}. If this is the case, then the building blocks
of life may have been present billions of years before the
formation of the Earth.

\section{Conclusion}
We report three direct spectroscopic detections of the \bump\ dust
absorption feature in quasar absorption systems at redshifts $1.4
\lesssim z \lesssim 1.5$. These are the first detections of this
feature in individual \mgii\ absorption systems.  From fitting the
composite SDSS quasar spectra reddened by a CCM Galactic
extinction law to the observed spectra, we derive best-fit
reddening parameters $E(B-V) \approx 0.14$, $R_V \approx 0.7$ for
SDSS J1446+0351, $E(B-V) \approx 0.13$, $R_V \approx 1.9$ for SDSS
J1459+0024 and $E(B-V) \approx 0.23$, $R_V \approx 5.5$ for SDSS
J0121+0027. The various $R_V$ values in these systems compared to
the average Galactic value of $R_V = 3.1$ indicates a wide range
of dominant grain sizes among intervening absorption systems. We
found that although the intrinsic slope of the quasar continuum in
each system is unknown, we can still meaningfully constrain the
$R_V$ value in each system. If the presently favored PAH model for
the \bump\ feature carrier is correct, we have detected complex
organic molecules in the young universe, about 9 billion years
ago.  The presence of PAHs at such large lookback times may be of
astrobiological interest.

\acknowledgements
J. W. and J. G. acknowledge support from NSF grant AST-01-38235, AST-02-43090 and NASA grants NAG 5-12115,
and NAG 5-11427. P. B. H. acknowledges support from the Department of
Astrophysical Sciences at Princeton University.
D. P. S. acknowledges NSF grant NSF03-007582.  We thank the referee for his useful comments and suggestions. 
Funding for the creation and distribution of the SDSS Archive has been provided
by the Alfred P. Sloan Foundation, the Participating Institutions, the National
Aeronautics and Space Administration, the National Science Foundation, the U.S.
Department of Energy, the Japanese Monbukagakusho, and the Max Planck Society.
The SDSS Web site is http://www.sdss.org/.  The SDSS is managed by the
Astrophysical Research Consortium (ARC) for the Participating Institutions.
The Participating Institutions are The University of Chicago, Fermilab, the
Institute for Advanced Study, the Japan Participation Group, The Johns Hopkins
University, Los Alamos National Laboratory, the Max-Planck-Institute for
Astronomy (MPIA), the Max-Planck-Institute for Astrophysics (MPA),
New Mexico State University, University of Pittsburgh, Princeton University, the United States Naval Observatory,
and the University of Washington.


\tabletypesize{\scriptsize}
\begin{deluxetable}{llccrc}
\tablecaption{Strong lines identified in dusty intervening
absorption systems} \tablehead{$\lambda_{\rm rest}$(\AA) & Ion &
$\lambda_{\rm obs}$(\AA)  & $EW_{\rm rest}\pm 1 \sigma$ error(\AA)
& $f$ & $z$} \tablewidth{0pt}
\startdata
             &            &               &\bf{SDSS J1459+0024 $z_Q=3.0124$}              &             &                   \\
\hline \\
2852.9642     &       MgI     &   6830.6       &  0.50$\pm$0.04   &            1.81    &          1.3942\\
2803.531      &       MgII      &      6713.6  &    1.67$\pm$0.04   &          0.295    &    1.3947\\
2796.352       &      MgII   &          6695.7   &  1.96$\pm$0.04    &          0.592      &   1.3945\\
2600.1729        &    FeII      &   6226.4       &  0.92$\pm$0.08    &          0.203     &  1.3946      \\
2586.6500    &        FeII   &  6194.6           &  0.58$\pm$0.17  &            0.0573    &        1.3947\\
2382.7652      &      FeII     &        5705.6   &  1.09$\pm$0.12    &          0.328      &  1.3946\\
2344.2139      &      FeII     &       5613.0   &  0.50$\pm$0.04    &          0.108     &        1.3945\\
1670.7874       &     AlII      &        4003.8    & 4.97$\pm$0.88     &         1.88       &      1.3959\\
\hline \\
             &            &               &\bf{SDSS J1446+0351 $z_Q=1.9452$}              &             &                   \\
\hline \\
2803.531      &       MgII      &   7022.5      &   3.66$\pm$0.24   &          0.295    &         1.5113\\
2796.352       &      MgII   &     7041.8       &  4.50$\pm$0.20    &          0.592      &       1.5117\\
2600.1729      & FeII    & 6532.5 & 2.31$\pm$0.12 & 0.203 & 1.5120\\
1670.7874 & AlII & 4196.03  &0.96$\pm$0.12 & 1.88 &  1.5114\\
\hline \\
             &            &               &\bf{SDSS J0121+0027 $z_Q=2.2241$}              &         & \\
\hline \\
2852.9642     &       MgI     &         6814.0  &   0.92$\pm$0.08  &            1.81    &          1.3884\\
2803.531      &       MgII      &       6695.0  &    2.72$\pm$0.08   &          0.295    &         1.3881\\
2796.352       &      MgII   &          6677.4   &  3.52$\pm$0.08    &          0.592      &       1.3879\\
2600.1729        &    FeII      &       6209.2   &  1.84$\pm$0.13    &          0.203     &        1.3880\\
2586.6500    &        FeII   &          6176.2   &  0.54$\pm$0.04  &            0.0573    &        1.3877\\
2344.2139      &      FeII     &        5597.7   &  1.05$\pm$0.17    &          0.108     &        1.3879\\
2026.137       &      ZnII       &       4836.6   &  0.50$\pm$0.04     &         0.412      &       1.3871\\
1862.7895      &      AlIII    &         4448.8   &  0.50$\pm$0.08    &          0.268       &     1.3882\\
1854.7164       &     AlIII     &        4430.2  &   1.05$\pm$0.13     &         0.539       &     1.3886\\
1808.0126        &    SiII       &       4309.2  &   0.75$\pm$0.08     &         0.0055      &     1.3834\\
1670.7874       &     AlII      &        3989.6    & 1.13$\pm$0.08     &         1.88       &      1.3879\\
\enddata
\tablecomments{Vacuum wavelengths, oscillator strength $f$ are from Morton (1991). Equivalent width is measured in the absorber restframe.}
\end{deluxetable}

\begin{deluxetable}{lccc}
\tablecaption{Fitting log with CCM extinction curve using different composite spectra}
\tablehead{Composite used & [$R_V$, $E(B-V)$] & Reduced \chisq\ }
\tablewidth{0pt}
\startdata
             &\bf{SDSS J1459+0024} &                   \\
\hline \\
Reddest & [1.9, 0.13] & 1.8 \\
Average & [0.7, 0.14] & 2.0 \\
Bluest & [0.7, 0.14] & 2.0 \\
\hline \\
             &\bf{SDSS J1446+0351} &                   \\
\hline \\
Reddest & [0.7, 0.15] & 2.6 \\
Average & [0.5, 0.16] & 3.0 \\
Bluest & [0.6, 0.18] & 2.9 \\
\hline \\
             &\bf{SDSS J0121+0027} & \\
\hline \\
Reddest & [6.0, 0.17] & 5.8 \\
Average & [6.0, 0.23] & 2.5 \\
Bluest  & [5.5, 0.23] & 1.3 \\
\enddata
\tablecomments{``Reddest'' and ``Bluest'' refer to the reddest and
bluest quartile composite quasar spectrum from Richards
et~al.\ (2003). ``Average'' refers to the composite quasar spectrum
using median combining from Vanden Berk et~al.\ (2001).}
\end{deluxetable}

\begin{deluxetable}{cccccc}
\tablecaption{Summary of the \bump\ measurements and metallicity estimates for three sightlines}
\tablehead{Line-of-sight & $z_{abs}$ & $\lambda_0$(\AA)\tablenotemark{a} & FWHM(\AA)\tablenotemark{b} & $N$(\hi) (cm$^{-2}$)\tablenotemark{c} & [X/H]\tablenotemark{d} }
\tablewidth{0pt}
\startdata
SDSS J1459+0024 & 1.3946 & 2230$\pm$20 & 400$\pm$60 &  $6.4\times 10^{20}$ & [Mg/H]\,$\simeq -2.7$ \\
SDSS J1446+0351 & 1.5115 & 2200$\pm$15 & 450$\pm$55 &  $6.9\times 10^{20}$ & [Mg/H]\,$\simeq -2.2$ \\
SDSS J0121+0027 & 1.3880 & 2250$\pm$25 & 400$\pm$60 &  $1.1\times 10^{21}$ & [Zn/H]\,$\simeq -0.2$; [Fe/H]\,$\simeq -2.6$\\
\enddata
\tablenotetext{a}{Central wavelength of the \bump\ feature.}
\tablenotetext{b}{Full Width Half Maximum of the \bump\ feature.}
\tablenotetext{c}{Estimated \hi\ column density in the absorption
system along line-of-sight.} \tablenotetext{d}{Estimated
metallicity [X/H]$\equiv \log$(X/H)$-\log($X/H)$_{\odot}$ in the absorption system. Reference solar abundances are from \citet{Sa96}.} 
\end{deluxetable}
\begin{figure}
\epsscale{1.00}
\plotone{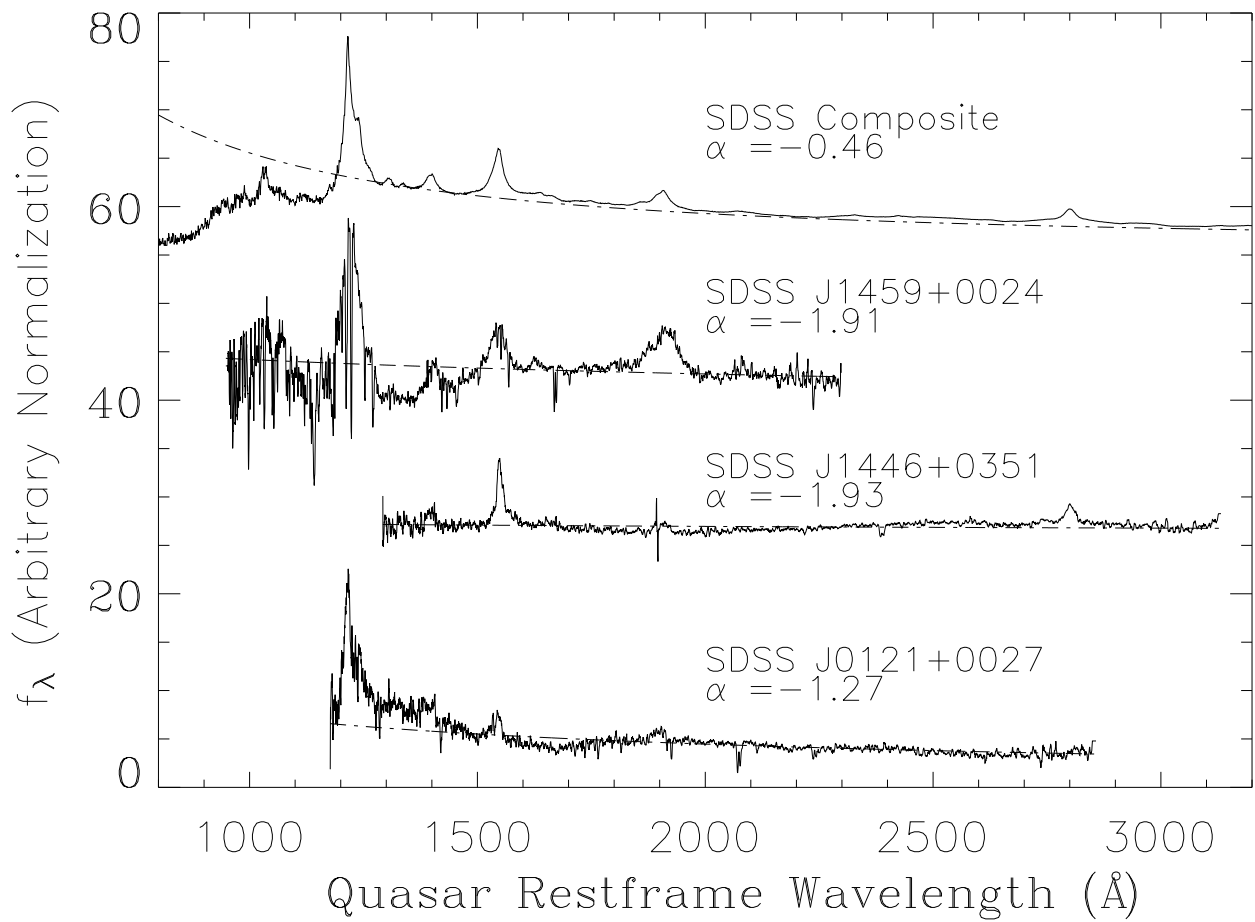}
\caption[]{ \singlespace Comparison between the composite quasar
spectrum from Vanden Berk et~al. (2001) (top; spectral index
$\alpha=-0.46$), the observed spectrum of SDSS J1459+0024 (second from top;
spectral index $\alpha=-1.91$), the observed spectrum of SDSS
J1446+0351 (third from top; spectral index $\alpha=-1.93$) and the observed spectrum of SDSS
J0121+0027 (bottom; spectral index $\alpha=-1.27$).
Power law fits to estimated continuum flux are shown as the dashed lines.
The broad absorption features at $\sim$1400~\AA\ in the quasar SDSS
J1459+0024 restframe, at $\sim$1900~\AA\ in the quasar SDSS
J1446+0351 restframe and at $\sim$1700~\AA\ in the quasar SDSS
J0121+0027 restframe are identified as the 2175~\AA\ dust
extinction feature.
}\label{figure1}
\end{figure}

\begin{figure}
\epsscale{0.6}
\plotone{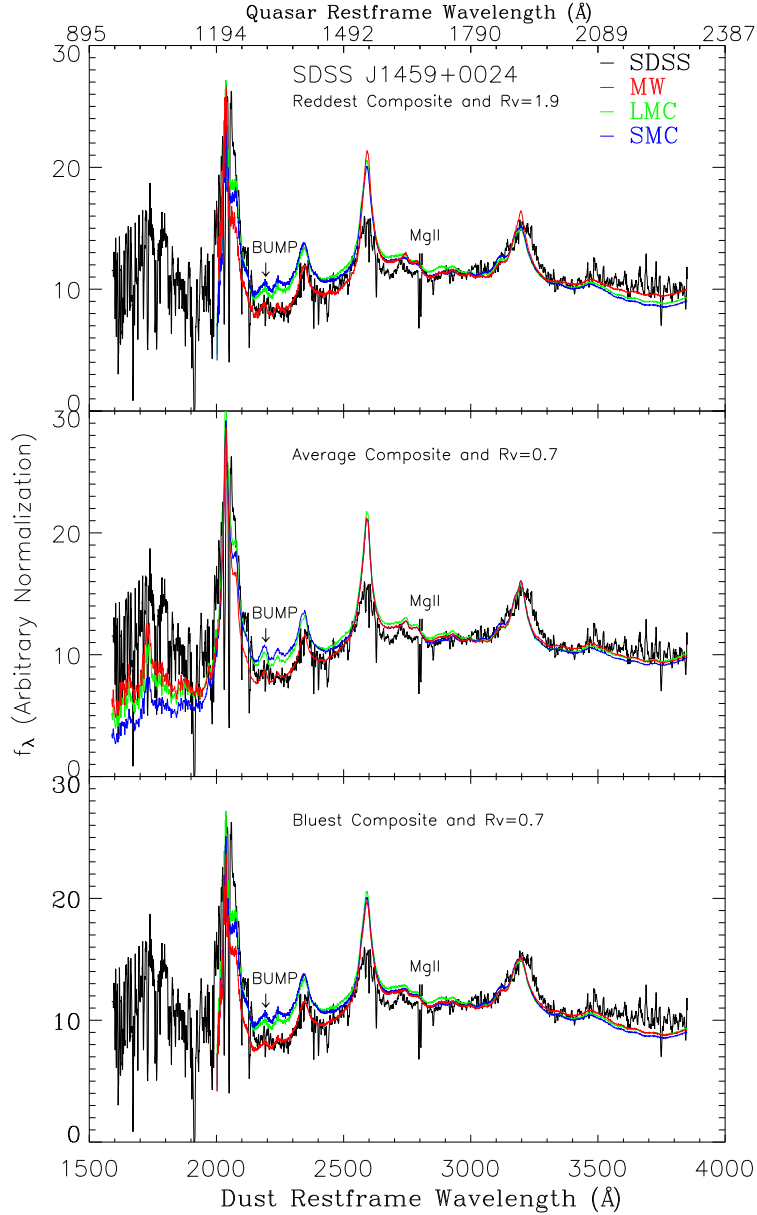}
\caption[]{\singlespace Comparison of the observed SDSS J1459+0024
spectrum (black) with model spectra obtained by reddening the SDSS
composite quasar spectrum (top panel: the reddest quartile
composite from \citet{Ri03}; middle panel: average composite from
Vanden Berk et~al. (2001); bottom panel: the bluest quartile
composite from \citet{Ri03}) with three types of extinction
curves: Milky Way with various $R_V$ (red), LMC (green), SMC
(blue). All spectra are normalized so that they have equal flux
density at 3200~\AA. The model spectra using the composite from
\citet{Ri03} can not cover the entire observed spectrum simply
because the bluest end of the composite starts from $\sim 1195$
\AA. The best fit is obtained using the reddest quartile
composite, reddened by a Milky Way-type extinction curve with
$R_V\approx 1.9$.} \label{figure2}
\end{figure}

\begin{figure}
\epsscale{0.6}
\plotone{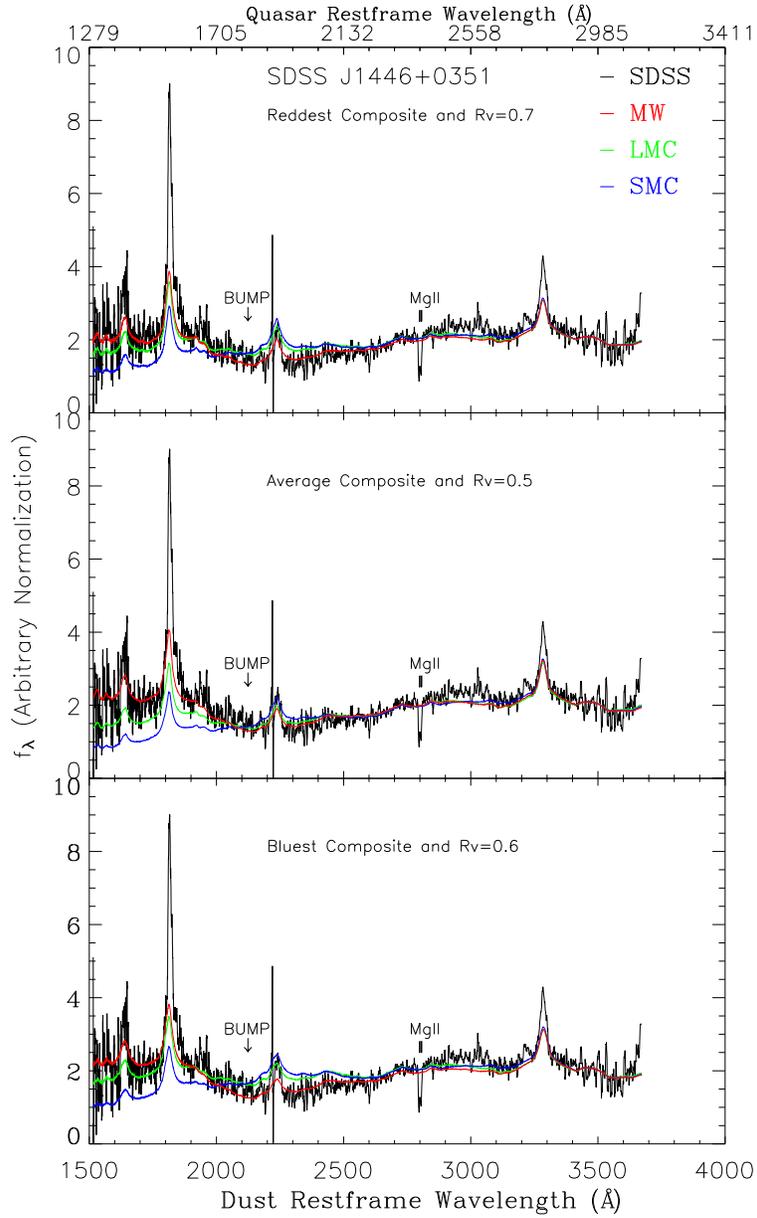}
\caption[]{\singlespace
Comparison of the observed SDSS J1446+0351 spectrum (black) with
model spectra same as Fig. 2 but for SDSS J1446+0351. All spectra
are normalized so that they have equal flux density at 3400~\AA.
The best fit is obtained using the reddest quartile composite,
reddened by a Milky Way-type extinction curve with $R_V\approx
0.7$.} \label{figure3}
\end{figure}

\begin{figure}
\epsscale{0.6}
\plotone{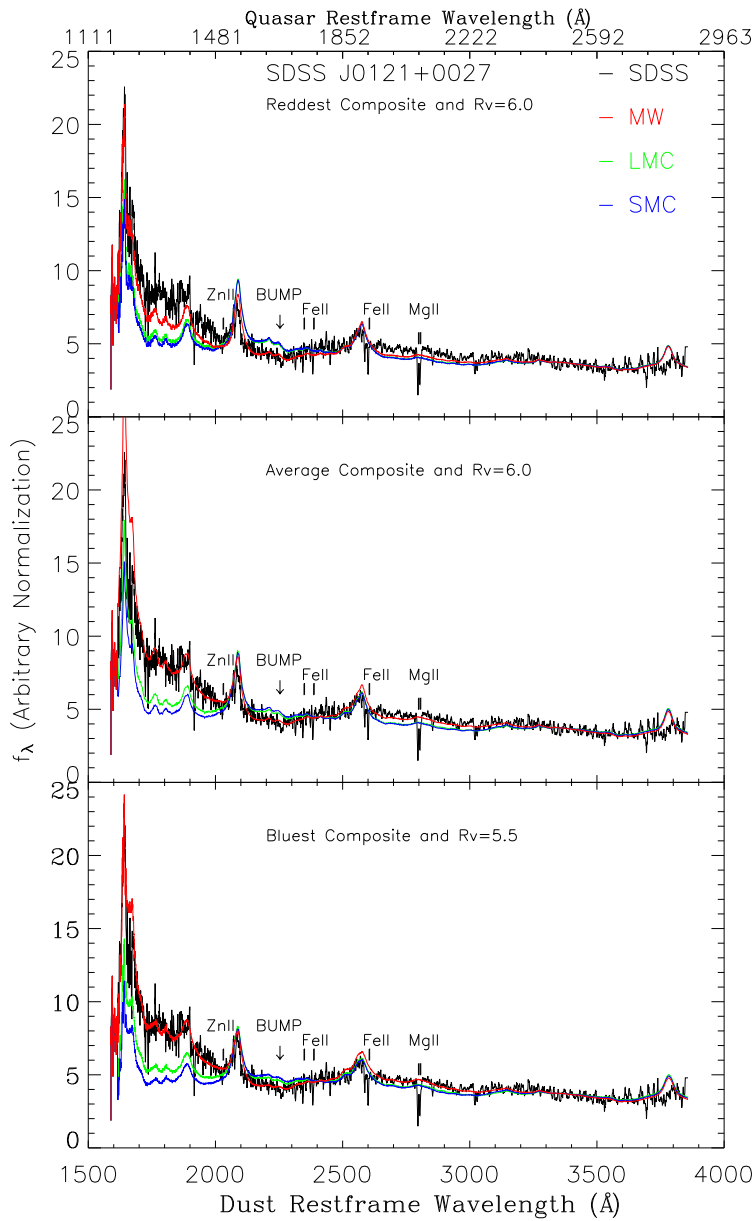}
\caption[]{\singlespace
Comparison of the observed SDSS J0121+0027 spectrum (black) with
model spectra same as Fig. 2 but for SDSS J0121+0027. All spectra
are normalized so that they have equal flux density at 3300~\AA.
The best fit is obtained using the bluest quartile composite,
reddened by a Milky Way-type extinction curve with $R_V\approx
5.5$.} \label{figure4}
\end{figure}

\end{document}